\documentclass{article}
\usepackage{graphicx} 
\usepackage[english]{babel}	
\usepackage{float}  

\usepackage{amssymb}
\usepackage{amsmath, amsfonts}
\usepackage{enumitem} 

\usepackage[left=2.5cm,right=2.5cm,top=2.5cm,bottom=2.5cm]{geometry}
\usepackage{xcolor}					

\usepackage[bookmarks,
            plainpages=false,
            pdfpagelabels,
            colorlinks=true,
            linkcolor=blue,
            anchorcolor=gray,
            citecolor=blue,
            urlcolor=gray,
            hyperindex=true,
            breaklinks]
            {hyperref}

\usepackage[authoryear]{natbib}

\usepackage{float}  
\usepackage{multirow} 
\usepackage{tikz}  
\usepackage{subcaption} 

\begin{document}
\begin{center}
{\Large
	{\sc  Multivariate LSTM Modeling for Wastewater Quality Prediction}
}
\bigskip

 Sophie Laruelle $^{1}$$^{2}$ \& Innocentia Bénédicta Sossou $^{1}$$^{3}$
\bigskip

{\it
$^{1}$ Univ Paris Est Creteil, Univ Gustave Eiffel, CNRS, LAMA UMR8050, F-94010 Creteil, France \\
$^{2}$ sophie.laruelle@u-pec.fr \\
$^{3}$ innocentia-benedicta.sossou@u-pec.fr }
\end{center}

\bigskip

{\bf Abstract:} The proposed approach relies on high-frequency multivariate time series collected at the Seine aval (SIAAP \footnote{Syndicat Interdépartemental pour l’Assainissement de l’Agglomération Parisienne}) treatment plant, including pH, temperature, conductivity, and total suspended solids, as well as exogenous variables related to precipitation and measurements from an upstream plant. The objective is to forecast wastewater quality over a 24-hour horizon. A multivariate Long Short-Term Memory (LSTM) recurrent neural network is implemented to capture complex temporal dependencies and nonlinear patterns in the data. The model is trained on one year of data and validated on four months of data (one per season), and is compared with a persistence model and SARIMA models. The evaluation shows an overall superiority of the LSTM model, particularly for variables exhibiting high levels of noise and nonlinearity.


\bigskip\bigskip

\section{Introduction}

\hspace{0.5cm} In multiple real-world applications, quantities are measured over time, giving rise to time series. A fundamental question is whether, based on past observations, it is possible to predict the future evolution of a series, either in the short or the long term. To address this issue, numerous studies have been conducted, leading to the development of models capable of representing the behavior of time series based on their main characteristics, such as trend, seasonality, and random fluctuations. Among the most prominent statistical models, the Autoregressive Moving Average (ARMA) framework, introduced by \citet{box1976analysis}, models a series by expressing present values as a function of past observations and error terms. Its extension to non-stationary time series exhibiting seasonal patterns has led to Seasonal Autoregressive Integrated Moving Average (SARIMA) models, which are widely used in practice, particularly in environmental and industrial applications \citep{effrosynidis2023time}. However, these models rely on linearity assumptions and are therefore limited to capturing nonlinear relationships between observations. In several real-world settings, time series exhibit complex dynamics and nonlinear relationships between variables. To overcome these limitations, machine learning approaches have been developed, particularly recurrent neural networks (RNNs), which are specifically designed to model sequential data. These models rely on an internal state that is continuously updated as new observations become available. Nevertheless, standard RNNs encounter difficulties when dealing with long input sequences, due to the loss of information over time, mainly due to the vanishing gradient problem \citep{bengio1994learning}. To address these limitations, Long Short-Term Memory (LSTM) architectures, a variant of RNNs introduced by \citet{hochreiter1997long}, have been proposed. These models incorporate both short-term and long-term memory mechanisms, enabling them to capture temporal dependencies over extended periods more effectively. They have since demonstrated strong performance in a wide range of applications, particularly for forecasting complex time series \citep{zhang2019deep}. The approach is based on a multivariate LSTM model applied to data describing wastewater quality at the inlet of the Seine aval (SIAAP) treatment plant. After presenting the context and the data, the configuration of the different models is detailed, followed by an analysis of the forecasting performance and a comparison of the approaches.

\section{Case study and datasets}

\begin{figure}[H]
    \centering
    \includegraphics[width=\textwidth]{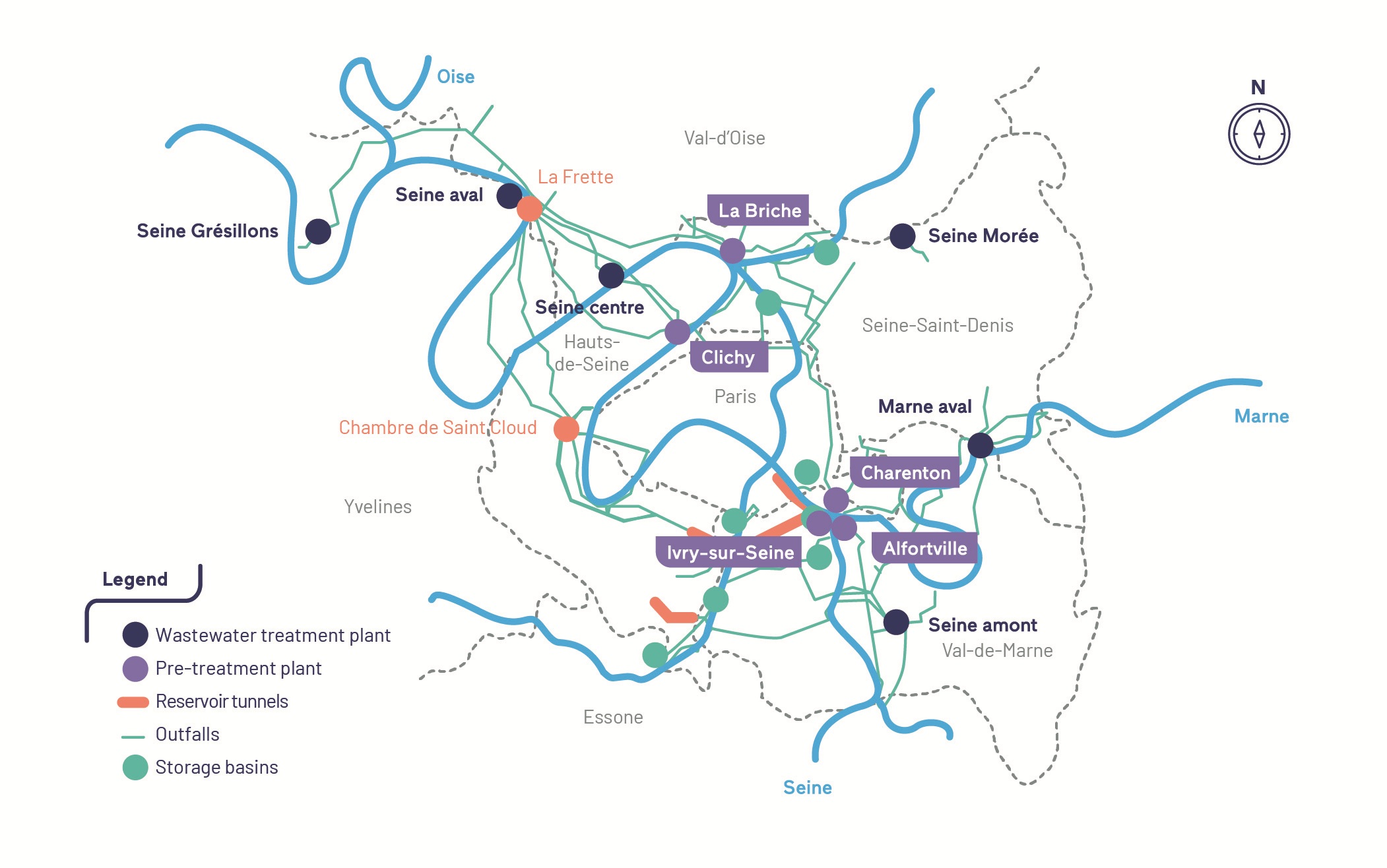}
    \caption{The wastewater treatment plants and the 5 outfalls arriving at Seine aval \protect\footnotemark.}
    \label{fig:reseau-siaap}
\end{figure}

\hspace{0.5cm} Efficient management of wastewater is a major challenge for large urban areas, both for environmental protection and for the optimization of treatment plant processes. In the Île-de-France region, SIAAP is responsible for the transport and treatment of wastewater, operating a network of approximately $400$ km (Figure \ref{fig:reseau-siaap}) of collectors and several treatment plants, including the Seine aval plant (La Frette), one of the largest in Europe.\\
\footnotetext{Picture from SIAAP -- \cite{innover_assainissement_2023}} 
In this context, the use of high-frequency sensor data, combined with additional information such as precipitation measurements and data from other treatment plant, enables the development of predictive models capable of capturing the multivariate dependencies and temporal dynamics of the series. The objective is to predict wastewater quality (based on the variables pH, temperature, conductivity and total suspended solids (TSS)) at the inlet to the Seine aval (SAV) wastewater treatment plant over a 24-hour time horizon, taking into account historical data, data from the Seine centre (SEC) plant, and precipitation measurements at both treatment plants.

Before reaching the Seine aval pre-treatment plant, wastewater from the five outfalls passes through La Frette, where both admitted and discharged flows into the Seine are measured. On arrival at the Seine aval pretreatment plant from La Frette, the water from the 5 outfalls is qualified (pH, conductivity, and oxidation-reduction potential measurements), then mixed. At the Seine aval plant, the following data are available at a frequency of 15 minutes: conductivity (in $\mu$S/cm), TSS (in mg/L), pH, temperature (in \textdegree C), and total flow (in m$^3$/s) from January 1, 2018, to December 31, 2024. At the Seine centre plant, pH, temperature, and turbidity (NTU\footnote{Nephelometric Turbidity Unit}) are available at a frequency of 5 minutes from November 25, 2022, to December 31, 2024. Total flow is also available at the Seine centre plant at a frequency of 5 minutes from January 1, 2022, to June 30, 2025. Precipitation measurements at both treatment plants are available at an hourly frequency from January 1, 2018, to August 18, 2025. All data were resampled to a uniform 15-minute frequency. Additionally, a 6-hour ahead forecast of total flow is available for both treatment plants. 

The data collected from Seine aval show missing data and wide variations, with peaks that will be difficult to predict, and which can be assumed to be outliers. The Seine centre data show greater variations than the Seine aval data (the sampling frequency is three times higher there). The dataset also contains missing values. Therefore, it is necessary to validate these data before using them in modeling. This analysis was conducted by \citet[see p. 174-179]{innover_assainissement_2023}. Using the statistical properties of these series and expert assessments, the authors identified coherent value ranges for the various analyzed parameters.


\section{Model configuration}

\hspace{0.5cm}In general, for time series forecasting, when future values are predicted from past observations, several linear models can be used as a first approximation. The simplest approach consists in assuming that the future value is identical to the most recently observed value (persistence model), or in resorting to autoregressive models that exploit past values of the target variable. These approaches rely on the assumption of a linear relationship between past and future observations \citep{box1976analysis}. However, several studies have shown that real-world time series may exhibit nonlinear dynamics in addition to these classical linear structures. This is the case in \citet{Gasparin2019DeepLF}, where electrical load forecasting is addressed. In this context, neural network-based models have been introduced to capture such nonlinear relationships. Recurrent neural networks, and in particular Long Short-Term Memory (LSTM) architectures \citep{hochreiter1997long}, are therefore particularly well suited for multivariate time series forecasting problems involving complex temporal dependencies.


\subsection{LSTM model and hyperparameter selection}

\hspace{0.5cm} A multivariate LSTM configuration is used to model and forecast the four variables of interest, namely pH, temperature, conductivity, and TSS. Prior to model training, a Min–Max scaling transformation was applied to normalize the data within the interval $(-1,1)$ over the training and validation sets. Input–output sequences were then constructed using a sliding window approach. An input sequence length of ten days (corresponding to 960 time steps at a 15-minute sampling rate) was used to predict the next $h$ time steps ahead (e.g., $h = 96$ corresponds to a one-day forecasting horizon). Following the model specification, a key issue concerns the selection of its hyperparameters, including the optimization algorithm, the learning rate, the loss function used as the objective function, the number of LSTM layers, and the number of units per layer. As recommended by \citet{geron2022hands}, a faster variant of stochastic gradient descent such as Adam (Adaptive moment estimation, \citet{kingma2015adam}) is particularly suitable for large-scale neural networks. Recent studies in real-world applications have also adopted Adam \citep{song2024multi, yaqoob2025predictive}, and it is therefore used in this work. The model is trained using the Huber loss function, which is more robust to outliers than the mean squared error. 
Finally, a Bayesian grid search procedure is employed to determine the optimal number of layers, the number of units per layer, and the learning rate. 

In summary, the proposed LSTM architecture is characterized by:
 
\begin{center}
\begin{tikzpicture}[scale=0.78, transform shape]

\draw[fill=blue!5, rounded corners=10pt, line width=1pt] (-7.6,-2.0) rectangle (7.6,2.0);

\node[align=left, font=\normalsize] at (0,0) {
\begin{tabular}{rl}
\textcolor{gray}{Input:} & \textbf{960 time steps and 12 features} \\
\textcolor{gray}{Output:} & \textbf{$h$ time steps} \\
\textcolor{gray}{Layers:} & \textbf{1 LSTM layer} \\
\textcolor{gray}{Hidden dimension:} & \textbf{256 units} \\
\textcolor{gray}{Optimizer:} & \textbf{Adam \citep{kingma2015adam}} \\
\textcolor{gray}{Stopping:} & \textbf{Early Stopping}
\end{tabular}
};

\end{tikzpicture}
\end{center}

\subsection{Baseline models}
\label{subsec:Baseline models}

\hspace{0.5cm} Two benchmark models are used in order to assess the contribution of the LSTM model to the modelling and forecasting of the studied variables.

\begin{itemize}

\item First, a persistence model is considered. For a forecasting horizon $h$, the model consists of repeating the observed trajectory over the previous $h$ time steps. In other words, the predicted sequence is given by:
$$
\hat{y}_{t+i} = y_{t+i-h}, \quad \text{for } i \in \{1, \dots, h\}.
$$
For instance, when $h = 96$ (corresponding to a $24$-hour horizon), the model predicts the next day by reproducing exactly the values observed during the previous day.

\item Second, a univariate SARIMA model is fitted for each target variable. A ten-day training window is used for model estimation, from November 27, 2023 at 8:00 p.m. to December 07, 2023 at 7:45 p.m. Both graphical analysis and a periodogram-based test reveal a daily seasonality, leading to $s=96$. The remaining hyperparameters are selected following the methodology of \citet{box1976analysis}. 

Table \ref{tab:sarima} summarizes the selected parameters for each variable. Further details on the model identification procedure are provided in Appendix~\ref{app:SARIMA identification}. These models serve as linear and naive benchmarks against which the nonlinear LSTM model is evaluated.
\begin{table}[H]
    \centering
    \begin{tabular}{|l|c |}
    \hline
     \textbf{Time series }   & \textbf{Models} \\
     \hline
        Conductivity SAV & SARIMA$(1,1,1)-(0,1,1)_{96}$ \\
        pH SAV  & SARIMA$(1,0,1)-(0,1,1)_{96}$ \\
        Temperature SAV  & SARIMA$(1,1,2)-(0,1,1)_{96}$ \\
        TSS SAV  & SARIMA$(1,0,0)-(2,1,0)_{96}$ \\
    \hline
    \end{tabular}
    \caption{Selected SARIMA hyperparameters for the target variables.}
    \label{tab:sarima}
\end{table}

\end{itemize}

\subsection{Training, validation, and test sets}

For LSTM training, one year of data from November 25, 2022, 07:00 a.m. to November 24, 2023, 06:45 a.m. was used. For validation, four months representing one month per season were selected: winter from November 24, 2023, 07:00 a.m. to December 23, 2023, 06:45 a.m.; spring from March 15, 2024, 12:00 a.m. to April 14, 2024, 11:45 p.m.; summer from July 1, 2024, 12:00 a.m. to July 31, 2024, 11:45 p.m; and fall from September 1, 2024, 12:00 a.m. to September 30, 2024, 11:45 p.m. Finally, for testing, one single day was used from December 07, 2023, 8:00 p.m. to December 08, 2023, 7:45 p.m. To compare the LSTM results on this test day, the SARIMA model was trained separately for each variable over a ten-day period immediately preceding the test set, from November 27, 2023, 08:00 a.m. to December 07, 2023, 07:45 p.m.

\begin{table}[ht]
\centering

\begin{tabular}{lllc}
\hline
Model & Dataset & Period & Number of observations \\
\hline
LSTM & Training &
Nov. 25, 2022 -- Nov. 24, 2023 &
$34944$ \\

 & Validation &
Seasonal months in 2023--2024 &
$11616$ \\

& Test &
Dec. 07--08, 2023 &
$96$ \\

SARIMA & Training &
Nov. 27, 2023 -- Dec. 7, 2023
&
$960$ \\

 & Test &
Dec. 07--08, 2023 &
$96$ \\
\hline
\end{tabular}
\caption{Datasets used for the different forecasting models.}
\label{tab:data_split}
\end{table}

\section{Forecasting results and model comparison}

\begin{figure}[H]
    \centering

    \begin{subfigure}{0.35\textwidth}
        \centering
        \includegraphics[width=\linewidth]{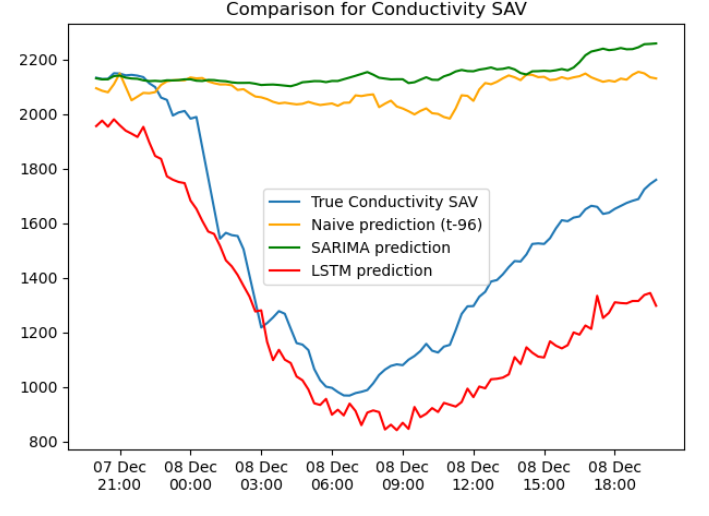}
    \end{subfigure}
    \begin{subfigure}{0.35\textwidth}
        \centering
        \includegraphics[width=\linewidth]{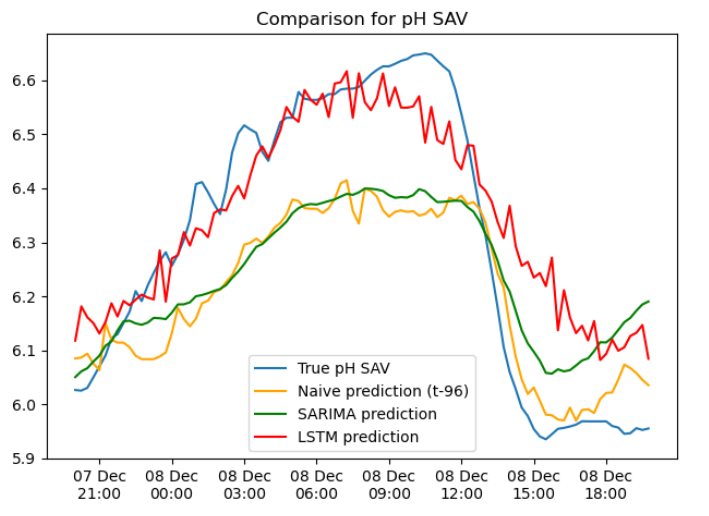}
    \end{subfigure}

    \begin{subfigure}{0.35\textwidth}
        \centering
        \includegraphics[width=\linewidth]{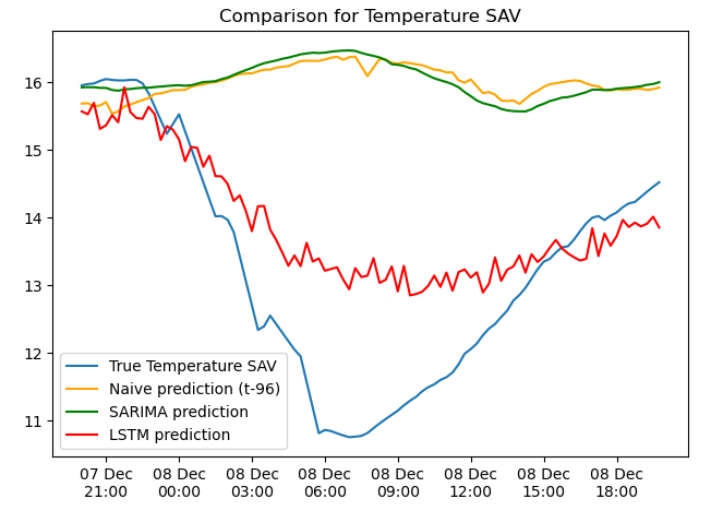}
    \end{subfigure}
    \begin{subfigure}{0.35\textwidth}
        \centering
        \includegraphics[width=\linewidth]{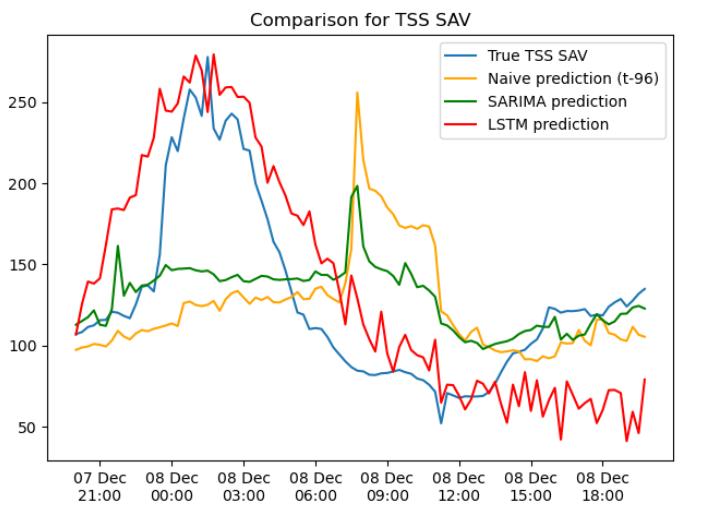}
    \end{subfigure}

    \caption{One-day forecasting comparison between the LSTM model and the baseline models.}
    \label{fig:predictions}
\end{figure}

\hspace{0.5cm} Model performance was evaluated using the Mean Absolute Error (MAE) and the Mean Absolute Percentage Error (MAPE), defined respectively as:
$$ \text{MAE} = \frac{1}{n} \sum_{t=1}^{n} |y_t - \hat{y}_t|,
\quad 
\text{MAPE} = \frac{100}{n} \sum_{t=1}^{n} \left| \frac{y_t - \hat{y}_t}{y_t} \right|.
$$
After training each model, predictions are generated for one day ($h = 96$).  
All models are evaluated on the same test set to ensure a fair comparison. The figures in Figure \ref{fig:predictions} illustrate the forecasting results for $h = 96$.

The performance measures for all variables are reported in Table~\ref{tab:results}. 

\begin{table}[H]
\centering
\small
\setlength{\tabcolsep}{2pt}

\begin{tabular}{l|cc|cc|cc}
\hline
\multirow{2}{*}{Variable} 
& \multicolumn{2}{c|}{Persistence} 
& \multicolumn{2}{c|}{SARIMA} 
& \multicolumn{2}{c}{LSTM} \\
\cline{2-7}
& MAE & MAPE (\%) & MAE & MAPE (\%) & MAE & MAPE (\%) \\
\hline

\multicolumn{7}{c}{\textbf{Horizon: $h = 96$}} \\
\hline
pH SAV & $$0.140$$ & $2.175^{\ast\ast}$ & $$0.153$$ & $$2.407$$ & $$0.097$$ & $1.579^{\ast}$ \\
Temperature SAV & $$2.872$$ & $$23.789$$ & $$2.826$$ & $23.488^{\ast\ast}$ & $$0.945$$ & $7.908^{\ast}$ \\
Conductivity SAV & $$602.056$$ & $48.176^{\ast\ast}$ & $$656.196$$ & $$52.611$$ & $$243.479$$ & $16.363^{\ast}$ \\
TSS SAV & $$49.902$$ & $$42.935$$ & $$38.291$$ & $32.679^{\ast\ast}$ & $$37.383$$ & $30.932^{\ast}$ \\
\hline

\end{tabular}

\caption{Forecasting performance (MAE and MAPE) of all models for different horizons and variable (\textit{$^{*}$ indicates the best performance (lowest error) and $^{**}$ indicates the second best performance for each metric within each row}).}
\label{tab:results}
\end{table}

The results indicate that the LSTM model outperforms both the persistence and SARIMA approaches for all variables on this test dataset. This test period was specifically selected to include strong variations compared to previous days, making it particularly challenging for forecasting models. The improvements are especially pronounced for highly non-linear and noisy variables such as Conductivity SAV and Temperature SAV, where the LSTM significantly reduces both error metrics. For TSS SAV, although the SARIMA model slightly outperforms the LSTM in terms of MAE, the LSTM remains competitive and achieves the best MAPE. The persistence model generally exhibits the poorest performance, particularly for variables with strong temporal variability, thereby confirming the added value of data-driven learning approaches.

\section{Conclusion and perspectives}

\hspace{0.5cm} In this study, LSTM neural networks were applied to multivariate time series forecasting in wastewater treatment plants. The results show that LSTM models are able to capture both short-term dynamics and long-term dependencies in high-frequency environmental data. Moreover, the LSTM model demonstrates competitive performance compared to classical SARIMA and persistence approaches, while offering greater robustness and stability. In particular, once the hyperparameters are tuned, the LSTM model can be applied to different test periods without the need for re-estimation, provided that the general data structure remains consistent. Nevertheless, in periods characterized by weak variability or predominantly linear dynamics, the LSTM model may perform similarly or slightly worse than simpler linear models. This suggests that no single model is universally optimal across all regimes. Future work should therefore focus on developing more stable and adaptive modelling strategies in terms of hyperparameter sensitivity. In particular, encoder–decoder architectures such as the LSTM-Attention-LSTM model \citet{wen2023time} have demonstrated improved performance over standard LSTM networks by integrating attention mechanisms to better capture temporal dependencies.

Beyond the Seine aval case study, the proposed workflow, from data preprocessing to 24-hour-ahead forecasting, can be applied to other wastewater treatment plants, provided that comparable data are available. In particular, this requires high-frequency measurements, a sufficiently long historical record, and multiple variables sampled at compatible frequencies. It is also desirable for the dataset to cover both dry periods and rainfall events, so that the model can learn the regime changes associated with these conditions. In this context, a multivariate LSTM model could be retrained using data from the new treatment plant. However, simple benchmark models, such as the persistence model or SARIMA, should first be evaluated before resorting to an LSTM model. Indeed, in the presence of strong seasonality, linear models may remain difficult to outperform when the series exhibit predominantly linear relationships and relatively stable dynamics. Conversely, the benefits of an LSTM model may be more pronounced in the presence of nonlinear relationships between variables, long-term temporal dependencies, or regime changes, for example between dry periods and rainfall events. Therefore, a practical implementation should begin by establishing simple benchmark models after data preprocessing. The temporal structure of the series should then be analyzed (ACF/PACF, seasonality and variability), and relevant explanatory variables should be identified before being incorporated into the model. Finally, the multivariate LSTM model could be retrained on the new data, and its performance gain relative to the benchmark models evaluated. However, the transferability of this approach to other wastewater treatment plants remains to be assessed.

\section*{Acknowledgments}

The authors would like to thank the Syndicat Interdépartemental pour l’Assainissement de l’Agglomération Parisienne (SIAAP) for its financial support through the Mocopée applied research program, as well as for providing the data used in this study. 
They also thank Jacques Printems for his support throughout this work.

\section*{AI Statement}

Generative AI tools were used during the preparation of this article to provide limited assistance with language and technical issues. These tools were mainly used to translate text from French to English and vice versa, improve grammar and wording, and help understand and correct errors in Python code and LaTeX documents. They were also occasionally used to generate small parts of code for specific tasks, which were reviewed and adapted by the authors. All scientific content, methodological choices, analyses, results, interpretations, and conclusions were developed and validated by the authors.

\bibliographystyle{apalike} 
\bibliography{references}

\appendix

\section{SARIMA hyperparameter identification}
\label{app:SARIMA identification}

It is well known that, if \( d \) and \( D \) are nonnegative integers, then the process \(\{X_t\}\) is a SARIMA\((p, d, q) -- (P, D, Q)_s\) process with period \( s \) if the differenced series \(
    Y_t = (1 - B)^d (1 - B^s)^D X_t
    \) follows a causal ARMA process given by  
    \begin{equation} \label{eq9}
    \phi(B) \Phi(B^s) Y_t = \theta(B) \Theta(B^s) \varepsilon_t, \quad \{\varepsilon_t\} \sim WN(0, \sigma^2),
    \end{equation}
    where  
    \(
    \phi(z) = 1 - \phi_1 z - \dots - \phi_p z^p, \quad \Phi(z) = 1 - \Phi_1 z - \dots - \Phi_P z^P, \quad \theta(z) = 1 + \theta_1 z + \dots + \theta_q z^q \text{ and } \Theta(z) = 1 + \Theta_1 z + \dots + \Theta_Q z^Q.
    \)\\
The seven hyperparameters ($p$, $d$, $q$, $P$, $D$, $Q$ and $s$) were determined for each univariate SARIMA model associated with the different variables.

\subsection{Checking for stationarity}

\begin{figure}[H]
    \centering
    \begin{subfigure}{0.35\textwidth}
        \centering
        \includegraphics[width=\textwidth]{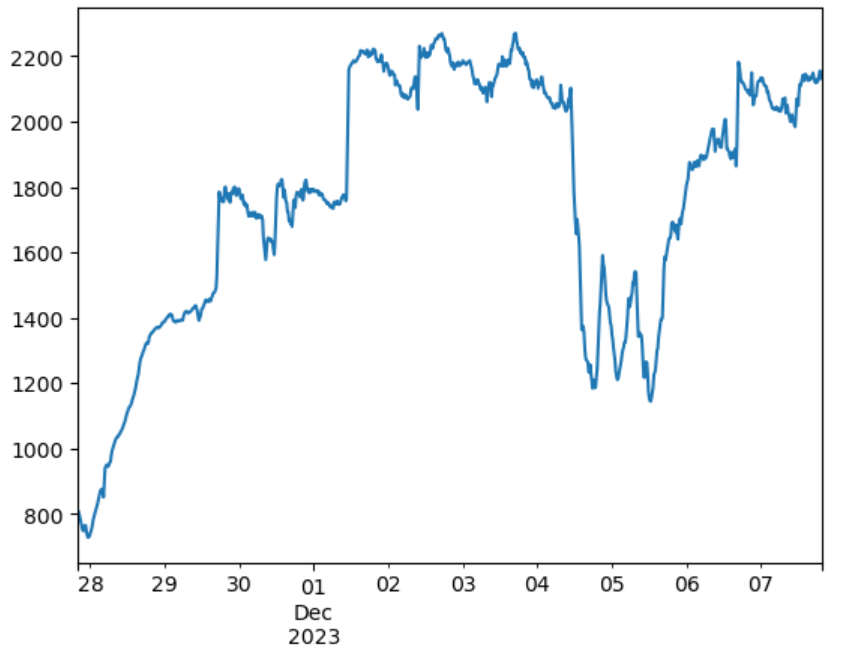}
        \subcaption{Conductivity SAV}
    \end{subfigure}
    \begin{subfigure}{0.35\textwidth}
        \centering
        \includegraphics[width=\textwidth]{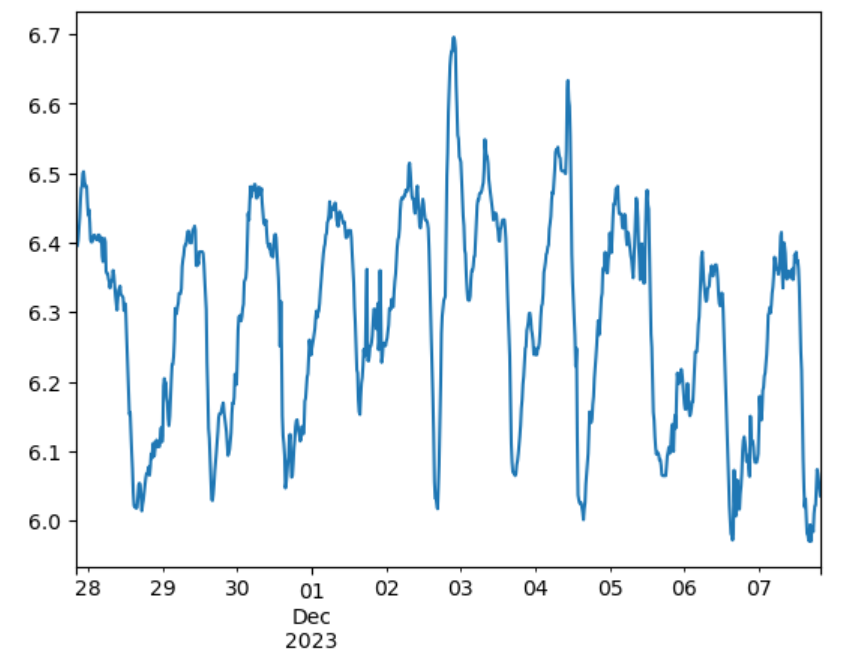}
        \subcaption{pH SAV}
    \end{subfigure}

    \begin{subfigure}{0.35\textwidth}
        \centering
        \includegraphics[width=\textwidth]{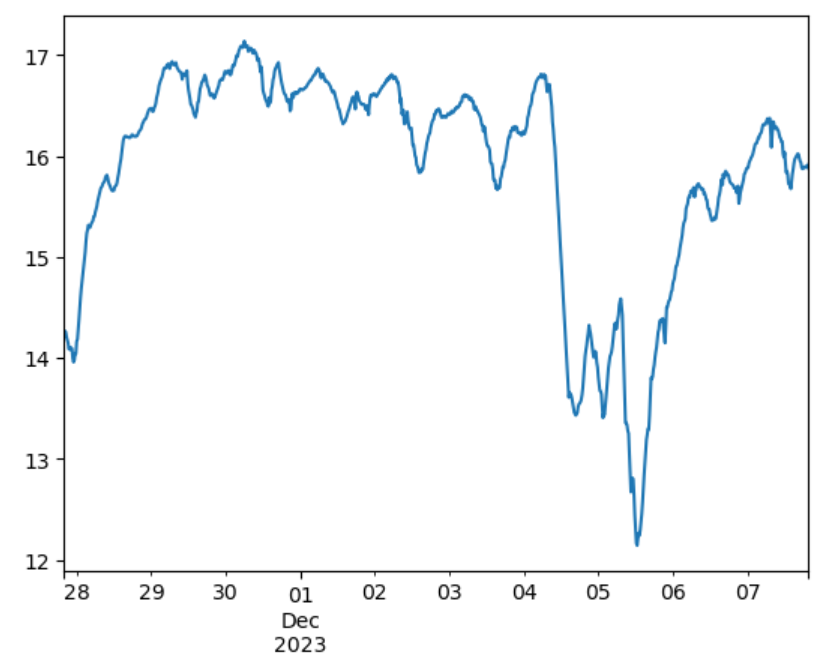}
        \subcaption{Temperature SAV}
    \end{subfigure}
    \begin{subfigure}{0.35\textwidth}
        \centering
        \includegraphics[width=\textwidth]{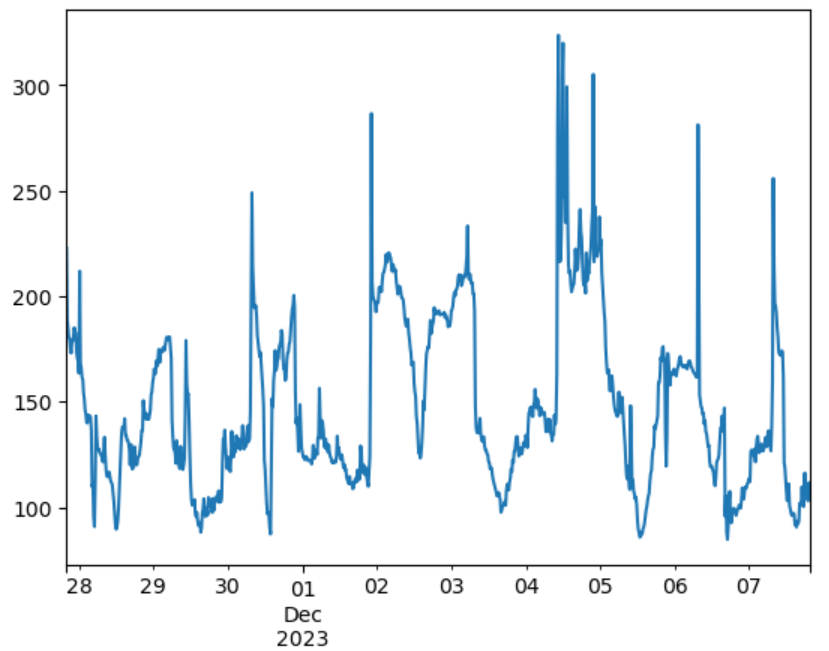}
        \subcaption{TSS SAV}
    \end{subfigure}

    \caption{Training dataset used for SARIMA training.}
    \label{fig:graph}
\end{figure}

Based on the graphical analysis of each series (Figure \ref{fig:graph}) and confirmed by the periodogram test, a seasonality equal to $s = 96$ was identified for all series. Then, $D=1$ was chosen for all time series.

To check the stationarity of the series for choosing each value of $d$, the Augmented Dickey–Fuller (ADF) test and the Kwiatkowski–Phillips–Schmidt–Shin (KPSS) test were used. The results presented in Table \ref{tab:stationnarity_test} were obtained using the 10-day training set.
 
\begin{table}[ht]
    \centering
    \begin{tabular}{|l|c|c|}
    \hline
       Differenciated time series (\( (1 - B^s) X_t \))  & ADF test & KPSS test \\
       \hline
       Conductivity SAV     & not stationary & not stationary \\
       pH SAV & stationary & stationary \\
       Temperature SAV & not stationary & not stationary \\
       TSS SAV & stationary & stationary \\
       \hline
    \end{tabular}
    \caption{Stationnarity testing results.}
    \label{tab:stationnarity_test}
\end{table}

Both tests produced similar results for the four differenced series on this training dataset. The pH SAV and TSS SAV series are stationary according to both tests; therefore, $d = 0$ and $D = 1$ for these series. In contrast, the Conductivity SAV and Temperature SAV series are not stationary. Therefore, the stationarity of the series $(1 - B)(1 - B^s) X_t$ was assessed, where $X_t$ denotes each time series (Table \ref{tab:stationnarity_test1}). 
\begin{table}[ht]
    \centering
    \begin{tabular}{|c|c|c|}
    \hline
       Differenciated time series (\( (1 - B)(1 - B^s) X_t \))  & ADF test & KPSS test \\
       \hline
       Conductivity SAV     & stationary & stationary \\
       Temperature SAV & stationary & stationary \\
       \hline
    \end{tabular}
    \caption{Testing for stationnarity.}
    \label{tab:stationnarity_test1}
\end{table}
Using both tests, these series were found to be stationary, implying that $d = 1$ and $D = 1$ for Conductivity SAV  and Temperature SAV. In general, Table \ref{tab:d_and_D} summarizes the values of $d$ and $D$ identified for all variables.
\begin{table}[h]
    \centering
    \begin{tabular}{|l|c|c|}
    \hline
      Time series &    $d$    &    $D$    \\
       \hline
       Conductivity SAV     & $1$ & $1$ \\
       pH SAV & $0$ & $1$ \\
       Temperature SAV & $1$ & $1$ \\
       TSS SAV & $0$ & $1$ \\
       \hline
    \end{tabular}
    \caption{$d$ and $D$ values.}
    \label{tab:d_and_D}
\end{table}

The parameter $d$ was allowed to vary between 0 and 1, as the test data may exhibit an unstable trend not observed in the training data.

\subsection{Models identification}

\begin{figure}[H]
    \centering

    \begin{subfigure}{\textwidth}
        \centering
        \includegraphics[width=\textwidth]{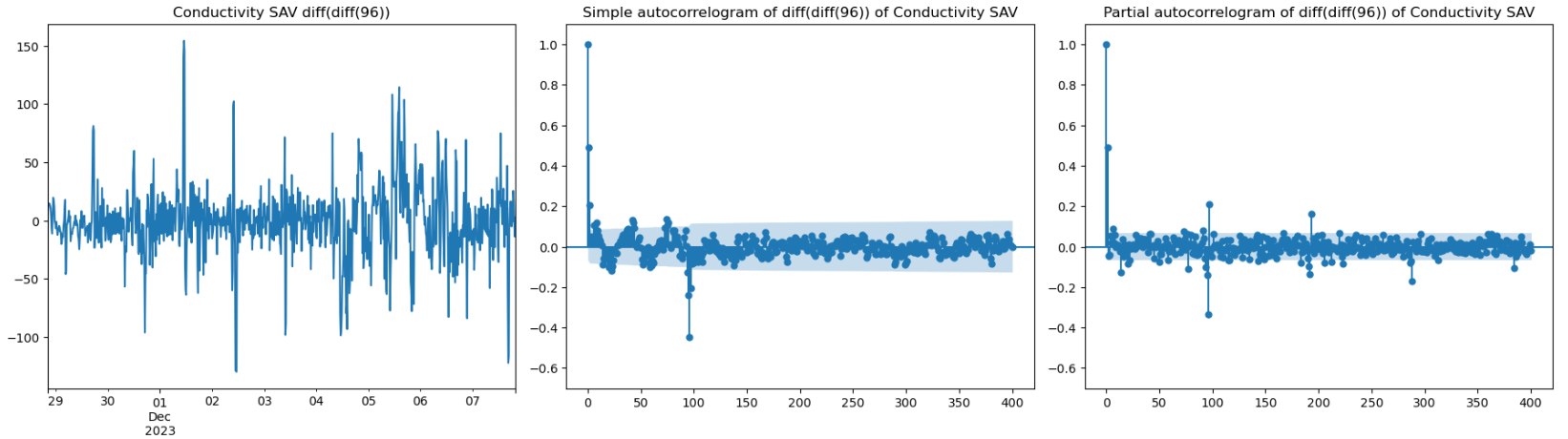}
    \end{subfigure}
    
    \begin{subfigure}{\textwidth}
        \centering
        \includegraphics[width=\textwidth]{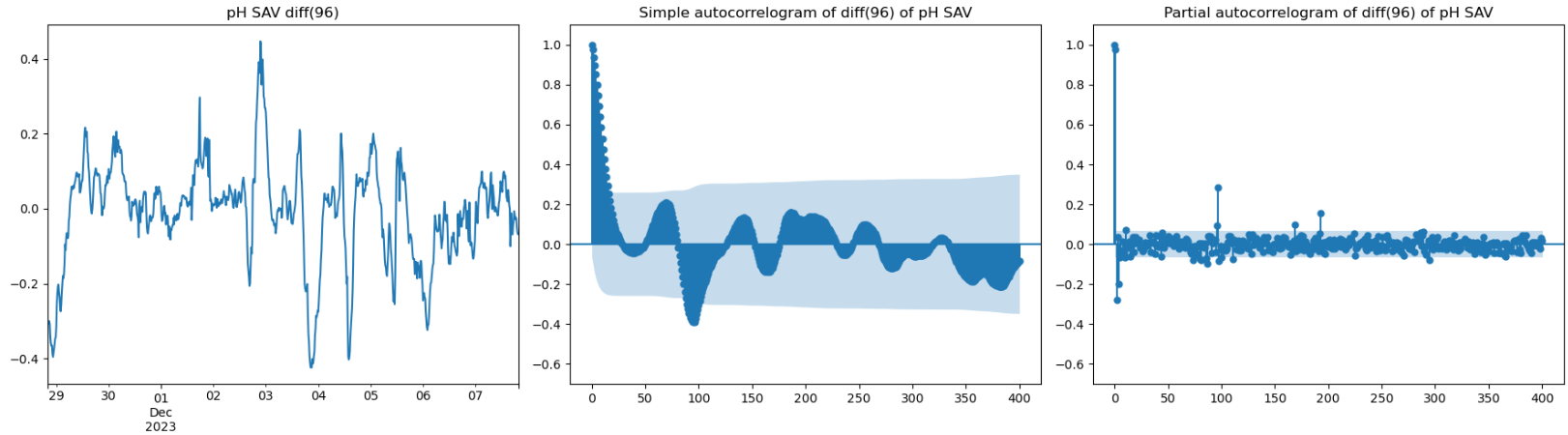}
    \end{subfigure}

    \begin{subfigure}{\textwidth}
        \centering
        \includegraphics[width=\textwidth]{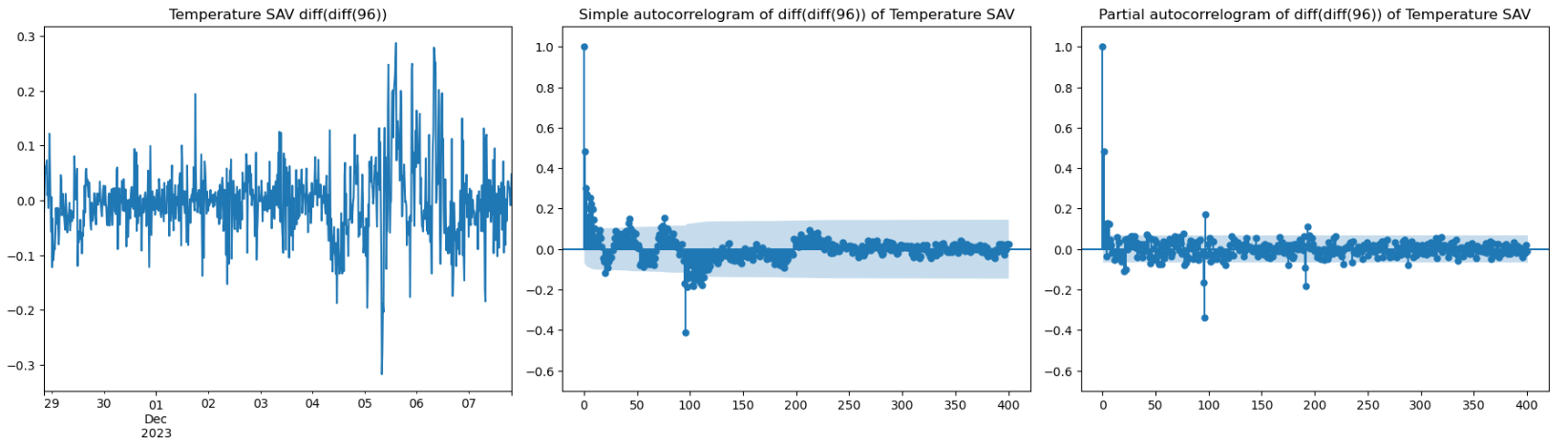}
    \end{subfigure}
    
    \begin{subfigure}{\textwidth}
        \centering
        \includegraphics[width=\textwidth]{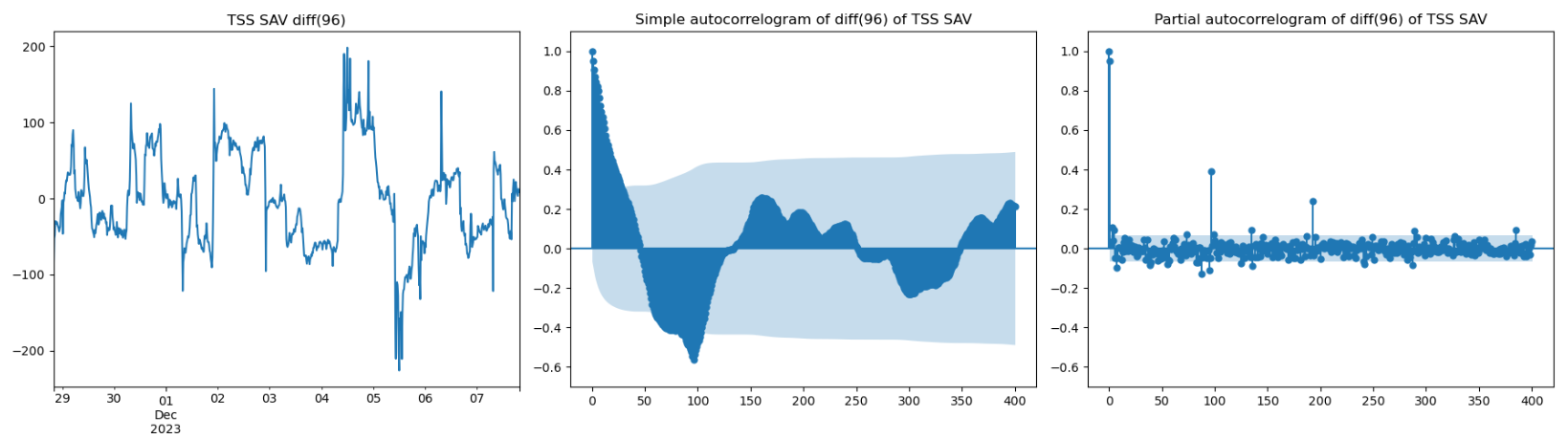}
    \end{subfigure}

    \caption{ACF and PACF plots used for identifying the parameters \(p\), \(q\), \(P\), and \(Q\).}
    \label{fig:acf_pacf}
\end{figure}

The next step is to determine the maximum value of the seasonal and non-seasonal orders of the AR components (parameters \( p \) and \( P \)) and MA components (parameters \( q \) and \( Q \)), based on the Autocorrelation function (ACF) and the Partial Autocorrelation function (PACF) plots of each stationary series. This may refer to the differenced series obtained after the previous step. The ACFs illustrate the correlation between a variable and its past values, while the PACFs evaluate the correlation between a variable and a specific lag, removing the effect of intermediate lags. Different maximum values of these parameters have been identified using the graphs in Figure \ref{fig:acf_pacf} and the Table \ref{tab:Maximum values param}.

\begin{table}
    \centering
    \begin{tabular}{|l|c|c|c|c|}
    \hline
     \textbf{Time series }   & $p_{max}$ & $q_{max}$ & $P_{max}$ & $Q_{max}$ \\
     \hline
        Conductivity SAV & 1 & 2 & 2 & 2 \\
        pH SAV  & 2 & 1 & 2 & 2 \\
        Temperature SAV  & 3 & 2 & 2 & 2 \\
        TSS SAV  & 2 & 1 & 2 & 2 \\
    \hline
    \end{tabular}
    \caption{Maximum values of $p$, $q$, $P$ and $Q$.}
    \label{tab:Maximum values param}
\end{table}

The appropriate values for the parameters $p$ and $q$ are indicated by the first significant autocorrelations outside the lower and upper confidence limits of the ACF and PACF graphs, while those for $P$ and $Q$ are indicated by the significant autocorrelations at lags multiples of $s = 96$ of the ACF and PACF graphs. However, it should be noted that in practice $P$ and $Q$ are typically less than three \citep{peter2016introduction}. For another example based on total flow in South Australia, see \citet{do2022wastewater}.

With the possible values of $p$, $P$, $q$ and $Q$, several configurations of these parameters were combined for each of the five variables, corresponding to $54$ potential models for Conductivity SAV, $36$ potential models for pH SAV, $81$ potential models for Temperature SAV and $36$ potential models for TSS SAV. The parameter $p$ was restricted to values starting from $1$, unlike the other parameters, which vary from $0$, as the ACF and PACF plots indicate the presence of an autoregressive component in all series. To identify the optimal model, all possible combinations of parameters \((p, d, q) - (P, D, Q, s)\) were evaluated. The selection of the best model for each series was based on several criteria: the Akaike Information Criterion (AIC), the Bayesian Information Criterion (BIC), and the Hannan–Quinn Information Criterion (HQIC), which balance model fit and complexity; the log-likelihood, which measures the goodness of fit; the convergence of the estimation algorithm; and the statistical significance of the estimated coefficients, assessed through p-values. The selected models are reported in Table \ref{tab:sarima} in Section~\ref{subsec:Baseline models}.

\subsection{Models diagnostic}

\begin{figure}[H]
    \centering

    \begin{subfigure}{\textwidth}
        \centering
        \includegraphics[width=\textwidth]{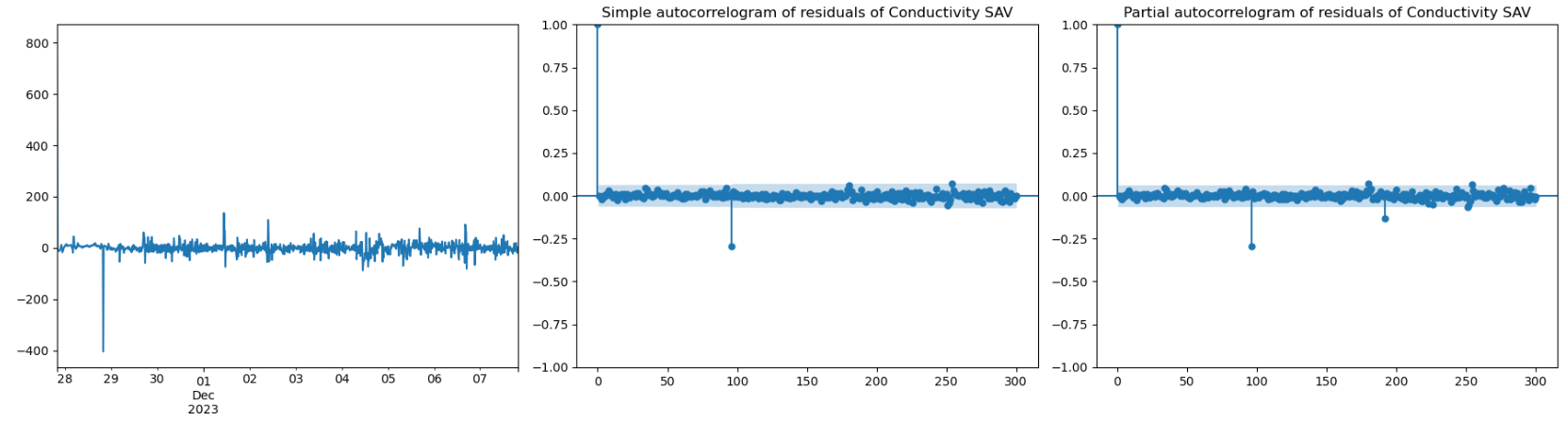}
    \end{subfigure}
    
    \begin{subfigure}{\textwidth}
        \centering
        \includegraphics[width=\textwidth]{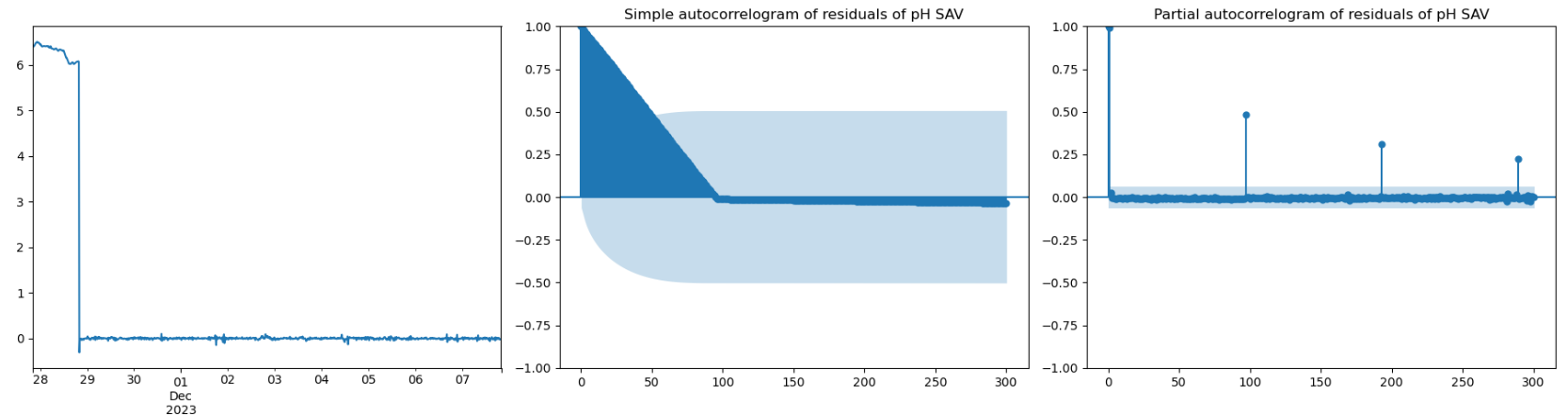}
    \end{subfigure}

    \begin{subfigure}{\textwidth}
        \centering
        \includegraphics[width=\textwidth]{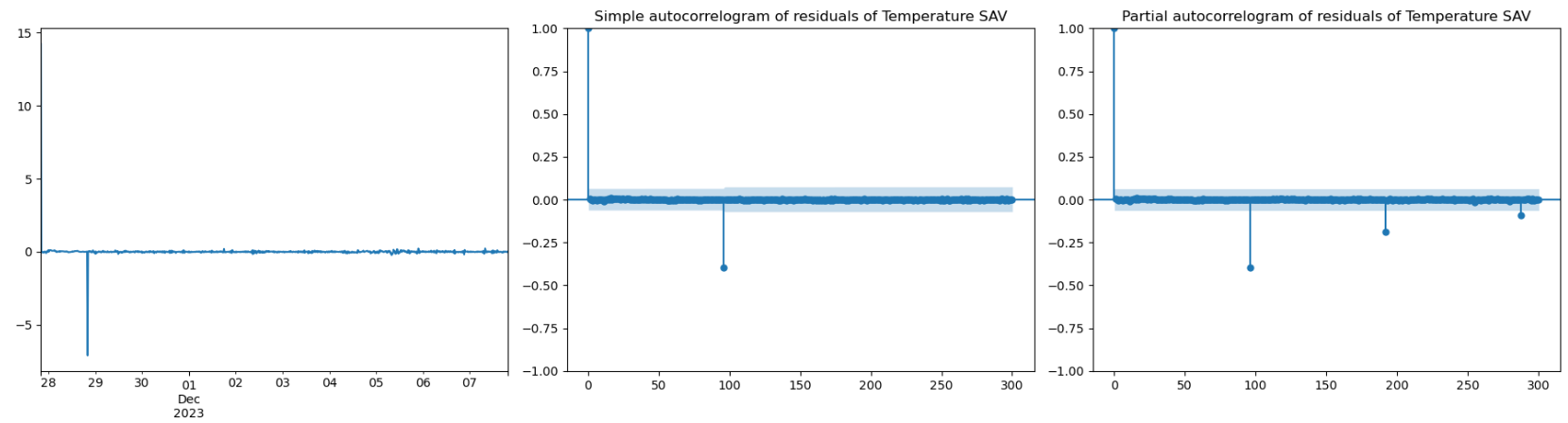}
    \end{subfigure}
    
    \begin{subfigure}{\textwidth}
        \centering
        \includegraphics[width=\textwidth]{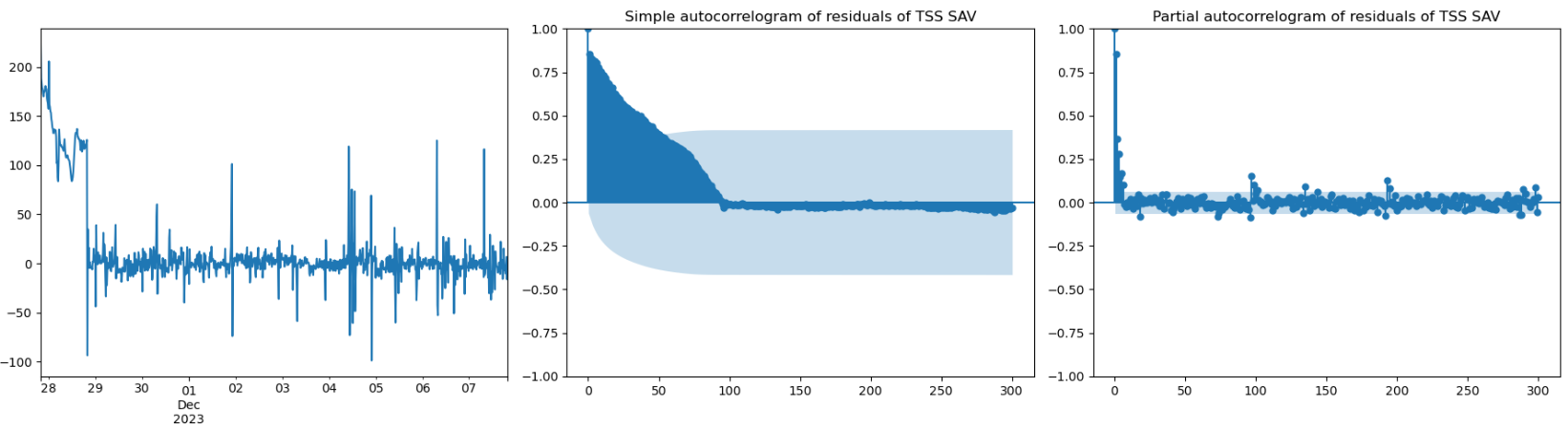}
    \end{subfigure}

    \caption{Residuals ACF and PACF.}
    \label{fig:acf_pacf_residuals}
\end{figure}

The Ljung-Box test indicated that residuals from models with better test data predictions were non-white noise ($\text{p-value} < 0.05$), while those predicting mean values were white noise. This is confirmed by the ACF and PACF plots of the residuals for each series in Figure \ref{fig:acf_pacf_residuals}. 

\begin{figure}[H]
    \centering
    \begin{subfigure}{0.35\textwidth}
        \centering
        \includegraphics[width=\textwidth]{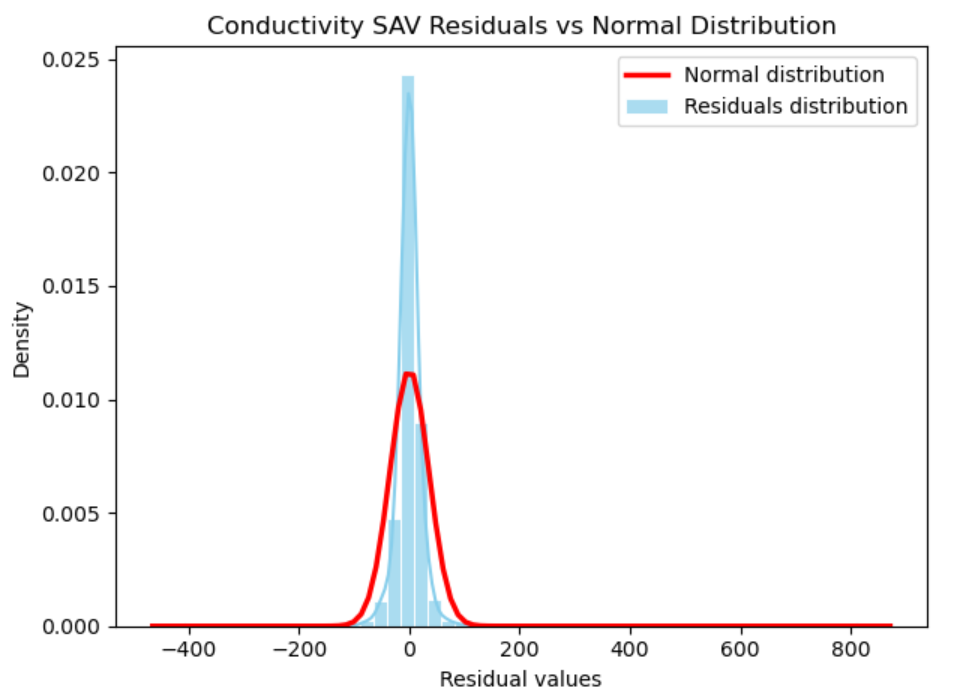}
    \end{subfigure}
    \begin{subfigure}{0.35\textwidth}
        \centering
        \includegraphics[width=\textwidth]{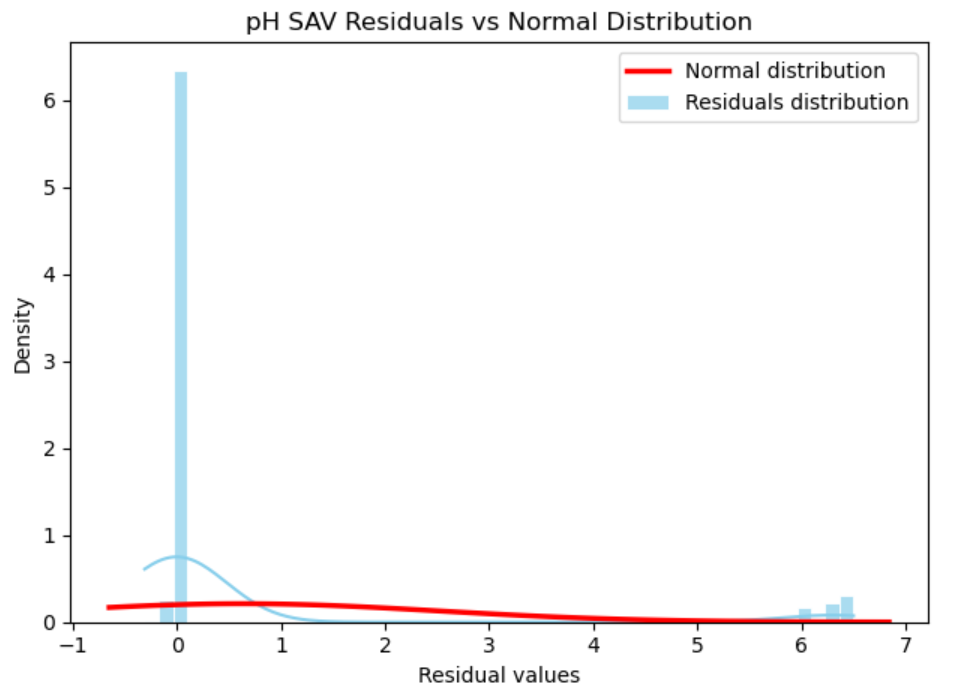}
    \end{subfigure}

    \begin{subfigure}{0.35\textwidth}
        \centering
        \includegraphics[width=\textwidth]{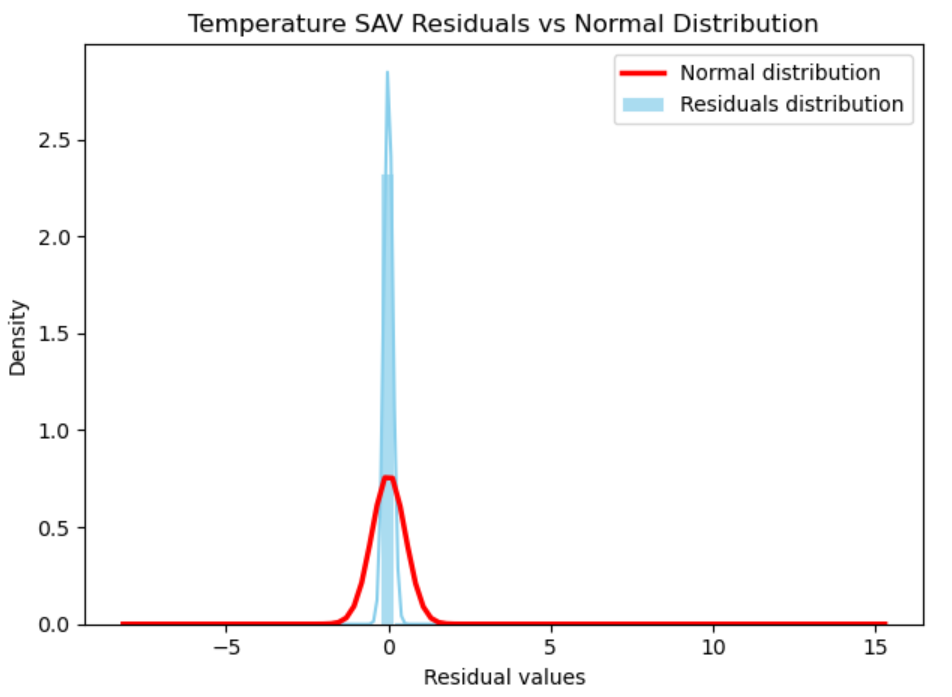}
    \end{subfigure}
    \begin{subfigure}{0.35\textwidth}
        \centering
        \includegraphics[width=\textwidth]{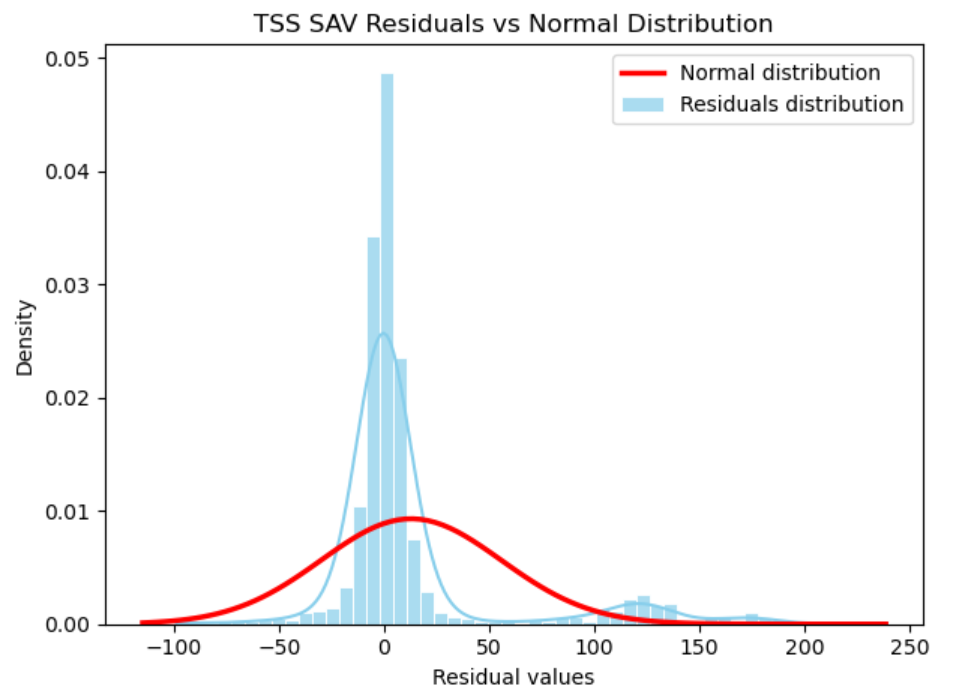}
    \end{subfigure}

    \caption{SARIMA models residuals vs Normal distribution.}
    \label{fig:densities}
\end{figure}

The residual histograms in Figure \ref{fig:densities} further indicate departures from normality. Although the residuals did not satisfy the white-noise assumption, the associated models were retained owing to their superior forecasting accuracy.

\end{document}